\documentclass[sigconf]{acmart}
\setcopyright{none}
\AtBeginDocument{%
  \providecommand\BibTeX{{%
    \normalfont B\kern-0.5em{\scshape i\kern-0.25em b}\kern-0.8em\TeX}}}

\setcopyright{acmlicensed}
\copyrightyear{2018}
\acmYear{2018}
\acmDOI{XXXXXXX.XXXXXXX}

\acmConference[Conference acronym 'XX]{Make sure to enter the correct
  conference title from your rights confirmation emai}{June 03--05,
  2018}{Woodstock, NY}
\acmBooktitle{Woodstock '18: ACM Symposium on Neural Gaze Detection,
 June 03--05, 2018, Woodstock, NY} 
\acmISBN{978-1-4503-XXXX-X/18/06}
\usepackage{xcolor}
\usepackage{tcolorbox}
\usepackage{amsmath}
\usepackage{amsthm}
\usepackage{amsfonts} 
\usepackage{mathrsfs}
\usepackage{fixmath}
\usepackage{mathtools}
\usepackage{booktabs}
\usepackage{balance}
\usepackage{subfigure}
\usepackage{xspace}
\newcommand{\method}{\texttt{LLM-SRR}\xspace}
\newcommand{\METC}{METC\xspace}

\begin{document}

\title{LLM-Powered Explanations: Unraveling Recommendations Through Subgraph Reasoning}


 \author{Guangsi Shi}
 \email{guangsi.shi@monash.edu}
 \affiliation{%
   \institution{Monash University}
   \streetaddress{Wellington Rd}
   \city{Clayton}
  \country{Australia}
   \postcode{3800}
 }

 \author{Xiaofeng Deng}
 \email{xiaofeng.deng43@gmail.com}
 \affiliation{%
   \institution{Shanghai University of Finance and Economics}
   \streetaddress{750, Dingguo street}
   \city{Shanghai}
   \country{China}
   \postcode{200433}
 }

 \author{Linhao Luo}
\email{linhao.luo@monash.edu}
\affiliation{%
	\institution{Monash University}
	\streetaddress{Wellington Rd}
	\city{Clayton}
	\country{Australia}
	\postcode{3800}
}
 \author{Lijuan Xia}
 \email{lijuan.xia@cn.bosch.com}
 \affiliation{%
   \institution{Bosch Corporate Research}
   \streetaddress{333, Fuquan street}
   \city{Shanghai}
   \country{China}
   \postcode{200050}
 }

 \author{Lei Bao}
 \email{lei.bao2@cn.bosch.com}
 \affiliation{%
   \institution{Bosch Corporate Research}
   \streetaddress{333, Fuquan street}
   \city{Shanghai}
   \country{China}
   \postcode{200050}
 }

 \author{Bei Ye}
\email{loretta.ye@cn.bosch.com}
\affiliation{%
	\institution{Bosch(China) Investment Ltd.}
	\streetaddress{333, Fuquan street}
	\city{Shanghai}
	\country{China}
	\postcode{200050}
}

 \author{Fei Du}
\email{rick.du@bshg.com}
\affiliation{%
	\institution{BSH Home Appliances Holding (China) Co. Ltd}
	\streetaddress{18, Qingjiang South Street}
	\city{Nanjing}
	\country{China}
	\postcode{210019}
}

 \author{Shirui Pan}
 \email{s.pan@griffith.edu.au}
 \affiliation{%
   \institution{Griffith University}
   \streetaddress{South Port}
   \city{QLD}
   \country{Australia}
   \postcode{4215}
 }

 \author{Yuxiao Li}
 \authornote{Corresponding author.}
 \email{yuxiao.li@cn.bosch.com}
 \affiliation{%
   \institution{Bosch Corporate Research}
   \streetaddress{333, Fuquan street}
   \city{Shanghai}
   \country{China}
   \postcode{200050}
 }

\newcommand{\LH}[1]{\textcolor{orange}{\small\textbf{[LH]} #1 $\triangleleft$}}
\renewcommand{\shortauthors}{Trovato and Tobin, et al.}

\begin{abstract}

Recommender systems (RecSys) are pivotal in enhancing user experiences across various web applications by analyzing the complicated relationships between users and items. Knowledge graphs (KGs), which model explicit relations between users and items, have been widely used to enhance the performance of recommender systems. However, a significant challenge persists in constructing KGs from unstructured data, such as reviews. Traditional information extraction tools fail to understand the complex subjective information inherent in the text such as preferences. Additionally, KGs are known to be noisy and incomplete, which are hard to provide reliable explanations for recommendation results.
An explainable recommender system is crucial for the product development and subsequent decision-making.
%
%
To address these challenges, we introduce a novel recommender that synergies Large Language Models (LLMs) and KGs to enhance the recommendation and provide interpretable results. 
Specifically, we first harness the power of LLMs to augment KG reconstruction. LLMs comprehend and decompose user reviews into new triples that are added into KGs. In this way, we can enrich KGs with explainable paths that express users' preferences. 
To enhance the recommendation on augmented KGs, we introduce a novel subgraph reasoning module that effectively measures the importance of nodes and discovers reasoning for recommendation. Finally, these reasoning paths are fed into the LLMs to generate interpretable explanations of the recommendation results.
Our approach significantly enhances both the effectiveness and interpretability of recommender systems, especially in cross-selling scenarios where traditional methods falter. The effectiveness of our approach has been rigorously tested on four open real-world datasets, with our methods demonstrating a superior performance over contemporary state-of-the-art techniques by an average improvement of 12\%. The application of our model in a multinational engineering and technology company (\METC)'s cross-selling recommendation system further underscores its practical utility and potential to redefine recommendation practices through improved accuracy and user trust.

\end{abstract}

\ccsdesc[500]{Do Not Use This Code~Generate the Correct Terms for Your Paper}
\ccsdesc[300]{Do Not Use This Code~Generate the Correct Terms for Your Paper}
\ccsdesc{Do Not Use This Code~Generate the Correct Terms for Your Paper}
\ccsdesc[100]{Do Not Use This Code~Generate the Correct Terms for Your Paper}

\keywords{Explainable Recommendation, Large Language Model, Knowledge Graph}


\received{20 February 2007}
\received[revised]{12 March 2009}
\received[accepted]{5 June 2009}

\maketitle

\section{Introduction}

\begin{figure}[thbp]
    \centering
    \includegraphics[width=8.5cm, height=7.5cm]{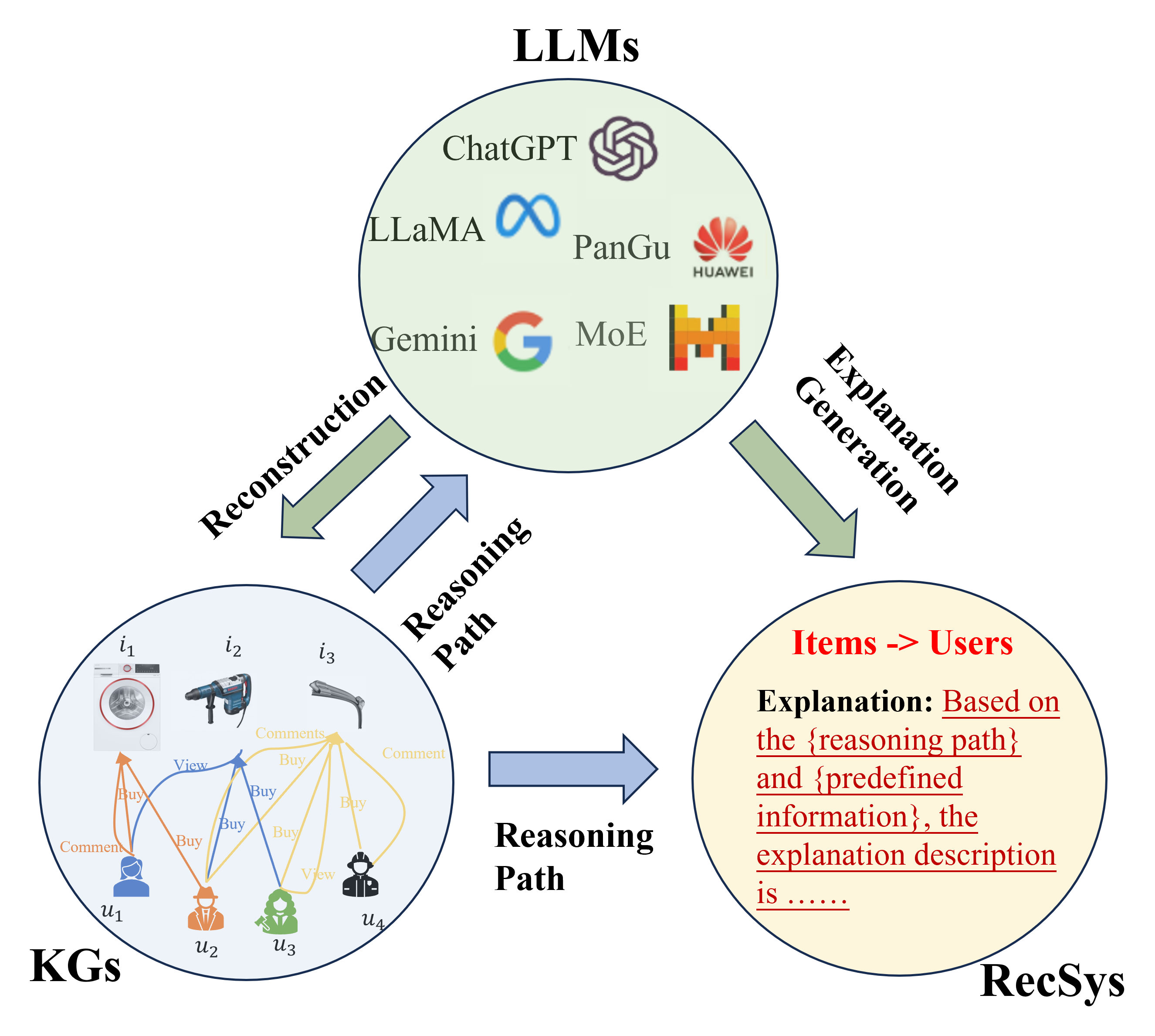}
    \vspace{-4mm}
    \caption{Main scenario and task: Arrows of different colors in KGs represent different channels (e-commerce platforms). Different products may not be sold in the same channel. The system needs cross channels to recommend and provide human-acceptable, reliable explanation descriptions.}
    \label{fig:overiew}
\end{figure}

Despite recommendation systems have become indispensable in mitigating information overload and enhancing user experience across modern web platforms and applications \cite{9416312}, most of recommendations fail to offer explanations, which is essential for decision-making by both users and e-commercial platforms. Consequently, there has been a paradigm shift in the recommendation system research community, with a growing emphasis not only on the accuracy of recommendations but also on their explainability. An explainable recommendation system significantly increases system transparency, boosts user trust and acceptance, and enhances recommendation efficiency \cite{balog2020measuring}.

Moreover, majority of literature focuses on natural interpretability of KG with rich structure reasoning \cite{DPMPN20} for improving recommendation system performance and explanation \cite{luo2020motif, zhang2024trustworthy}, but they often fail to construct the semantic relationships such as emotion, preference pertinent to users from review. This limitation hampers the interpretability, implicit relevance, and accuracy of recommendations derived from these models with noisy and incomplete information \cite{peng2023knowledge}. Extracting semantic information from text for knowledge graph construction faces several limitations. Unstructured data complexity and the absence of well-defined ontologies challenge traditional tools in relation extraction and entity linking. These tools often struggle with context understanding, leading to inaccuracies in entity-relation identification and semantic interpretation \cite{hofer2023construction}. Furthermore, one of the important way of KG reasoning for explainable recommender is rule-based methods \cite{PGPR19, ReMR22}, while these approaches can only depend on the existing paths and search an optimal path for explanation, and it does not work in the cross-selling scenarios, where the potential links may not be built in the original knowledge graph. It will result in "recommendation hallucination" which forces explanations solely for the sake of recommending. For example, different business units in multinational engineering and technology company (\METC) group operate their own channels for selling their products. To maximize production profits, it is essential to implement cross-selling across these units. Traditional strategies may not suffice for this purpose, underscoring the importance of an explainable recommendation system for effective cross-selling \footnote{Within a company group, cross-selling indicates that products from different business units/product lines of the company are successfully sold to one consumer.}, which is crucial for \METC's development. 

The advent of LLM, epitomized by advancements such as ChatGPT has attracted widespread attention and developed rapidly. Its excellent understanding capabilities and easy-powerful tools are capable of sophisticated reasoning and generation tasks in real-world applications \cite{pan2024unifying}. However, the way of successfully applying or combining their powerful capabilities into the recommendation system is still a promising but challenging task. 

According to the requirement and issues above, synergizing knowledge graphs with LLMs developments present a promising road map for recommender. However, the predominant mode of implict integration between knowledge graphs and LLMs remains embedding-based strategies \cite{wang2021kepler, ZHU2023119369}. This approach merges knowledge graph data embedding with word data embedding to inform the learning of user-item representations but often loses robust reasoning capabilities of LLMs. This deficiency is particularly pronounced when addressing diverse and specific business needs. Additionally, LLMs-enhanced KG systems \cite{luo2023chatrule,luo2024rog} with explicit modelling are constrained by their reliance on predefined meta-paths, necessitating a nuanced consideration of how to design effective rule paths tailored to specific business scenarios. More importantly, if fine-tuning LLMs for improving performance, it requires a lot of computing resources which is not friendly for actual marketing analysis.

In response to the identified challenges, we introduce a cutting-edge framework named \underline{\textbf{LLM}} powered \underline{\textbf{S}}ubgraph \underline{\textbf{R}}easoning to facilitate an explainable \underline{\textbf{R}}ecommendation system (\method for abbreviation)  which can be seen in Figure \ref{fig:overiew}. This novel architecture capitalizes on the unique strengths of LLM by prompt engineering and knowledge graphs to overcome the limitations of existing recommendation systems. The \method framework is delineated through a three-step process: \textbf{First, Information extraction and KG reconstruction by LLMs:} Our approach begins by extracting relevant information from user reviews. This process involves identifying new features based on predefined targets, which are critical in capturing the nuanced key works and semantic information of users by LLM's prompt engineering. A new KG integrates both original user-item relationships and the newly identified targets from user reviews. \textbf{Second, Subgraph Reasoning by Attendtion-based Diffusion Scoring:} This step involves the implementation of an attention mechanism for effective message passing in subgraph reasoning which skips out of the original connection pattern to search for more potential link relation, followed by a recommendation scoring process to rank the items. \textbf{Last, LLMs generating Explainable Description:} The LLM continues to leverage predefined keywords, the reasoning path generated by the subgraph, and the coherent descriptions acquired through thorough comprehension. Such explainable context aids front-end analysts in decision-making and planning processes. This method both enhances the recommendation's accuracy and ensures that the post-hoc explanation path and description behind each recommendation is transparent and understandable to both users and brands. In summary, the contribution of this paper is threefold:

\begin{itemize}
    
    \item \textit{Contribution 1:} To the best of our knowledge, this study represents the effort to a LLM powered explainable recommendation system by subgraph reasoning. This innovative alignment introduces a novel paradigm in the domain of personalized recommendation systems, especially in a novel cross selling scenario.
    
    \item \textit{Contribution 2:} We have developed customizable, user-friendly tools designed specifically for explainable recommender. These tools not only facilitate understanding the semantic information of user but also provide an autonomous post-hoc explanation description by LLMs, thereby enhancing transparency and understandability in recommender for marketing analyst.
    
    \item \textit{Contribution 3:} The efficacy of  subgraph reasoning module has been rigorously tested across three open source recommendation datasets, where it has demonstrated state-of-the-art performance. Furthermore, its applicability has been successfully validated in a real-world scenario involving cross-selling activities at METC, where it yielded highly favorable outcomes.

\end{itemize}

Therefore, our \method architecture effectively harnesses the capabilities of subgraph for reasoning  and the semantic understanding ability of LLMs to provide a highly explainable recommendation system. It addresses specific user and brand requirements through a visualizable and understandable recommendation pathway, significantly enhancing the persuasiveness of the system. Moreover, the nature of knowledge graphs, which can be updated based on unstructure information, alongside the system's ease of training, positions \method as a robust, adaptable, and user-centric solution in the realm of explainable recommendation systems.

\section{Related Works}
\subsection{Explainable Recommendation System}

Explainable recommendation systems have emerged as a pivotal enhancement in recommendation tasks, with their capacity to not only increase the efficacy of recommendations but also bolster their trustworthiness. Such systems are broadly categorized into two types: post-hoc and model-based explanations. In the realm of post-hoc explanation models, CountER \cite{CountER21} employs counterfactual constrained learning to derive succinct yet potent explanations for otherwise opaque recommendation models. Conversely, model-based explanations provide insights directly from the recommendation process itself. KPRN \cite{KPRN19} utilizes LSTM to process paths within KGs from users to items, generating embeddings for these paths which are then evaluated for their relevance. Similarly, EIUM \cite{EIUM19} focuses on explicating the semantic paths between users and items, thereby equipping the recommendation system with the capacity for path-wise explanation. RuleRec \cite{Rulerec19} introduces a rule-guided framework that derives rules from KGs for item recommendations. For a more personalized and explainable approach, PGPR \cite{PGPR19} and ReMR \cite{ReMR22} employ path reasoning and multi-level reasoning, respectively, through reinforcement learning to refine recommendations. Recent studies have begun to underscore the importance of subgraphs within knowledge graphs for enhancing explainability. GraIL \cite{GraIL20} pioneers inductive relationship predictions using subgraph reasoning. Moreover, GnnExplainer \cite{GnnExplainer19} and CF-GNNExplainer \cite{CF-GnnExplainer22} elucidate Graph Neural Networks (GNNs) via subgraph analyses. Within the specific context of recommender systems, GREASE \cite{GREASE22} innovatively employs subgraphs to furnish both factual and counterfactual explanations for GNN-based black-box models. However, most of them fall short of explicating the model's internal workings or enhancing model performance comprehensively and cannot make full use of other additional information to help reasoning such as text information and cannot skip original relation structures.

\subsection{Large Language Models and Knowledge Graphs Combination}

Researchers have explored integrating KG with LLM at different stages to enhance their capabilities. During pre-training, incorporating KG aids LLM in assimilating knowledge \cite{rosset2020knowledge}. At the inference stage, accessing KG bolsters LLM's performance in domain-specific knowledge \cite{NEURIPS2020_6b493230}. Furthermore, KG contribute to interpreting LLM by clarifying facts \cite{petroni2019language} and elucidating the reasoning process \cite{lin-etal-2019-kagnet}, thereby improving interpretability.

Moreover, knowledge graph often struggle with incompleteness \cite{bordes2013translating} and text corpus processing for KG construction \cite{zhu2023llms}. Leveraging LLM's generalizability, researchers are utilizing LLM to enhance KG tasks. By employing LLM as text encoders, they process textual content within KG, using the generated text representations to improve KG's comprehensiveness \cite{zhang2020pretrain}. Furthermore, LLM is applied to extract entities and relationships from text for KG creation \cite{kumar2020building}. Recent efforts focus on designing KG prompts that transform KG structures into formats understandable by LLM, facilitating direct LLM application in tasks like KG reasoning \cite{chen2023incorporating}.

The integration of KG and LLM have become a focal point of research, given their complementary nature  \cite{wang-etal-2021-kepler}. This synergy aims at creating a unified framework to leverage the strengths of both, enhancing their capabilities. The Synergized Model can enhance the mutual capabilities of LLM and KG, while the technique layer incorporates various methods to boost performance further. This integrated approach can be applied to real-world such as search engines \cite{thoppilan2022lamda}, recommender systems \cite{liu2023chatgpt}, and AI assistants \cite{sun2021ernie}, showcasing the practicality of our unified framework.

\subsection{Subgraph Reasoning}

Subgraph reasoning has emerged as a potent paradigm for enhancing the interpretability and performance of models across various domains and powerful ability of skip hop has been proved by \cite{feng2022powerful}. A novel approach \cite{han2020explainable} that combines temporal relational attention with reverse representation updates to guide subgraph extraction. In fake news detection, a reinforcement subgraph generation method \cite{yang2022reinforcement} alongside a hierarchical graph attention network improves both generalization and discrimination, offering clear explainability by identifying critical subgraphs. For fraud detection, SubGNN \cite{song2021subgraph}, leverages heterogeneous subgraphs and a relational graph isomorphism network for precise fraud identification without relying on global IDs.

A novel approach \cite{sun2023substructure} for inductive relation prediction incorporates substructure information into subgraph reasoning, significantly enhancing precision by utilizing semantic correlations between relations. Addressing scalability in KG, one-shot subgraph reasoning proposes a two-step prediction process that significantly increases efficiency and performance on large-scale KG \cite{zhou2023less}. For question answering, integrating subgraph-aware relation and direction reasoning into a novel neural model, RDAS \cite{wang2021integrating}, substantially improves answer precision by leveraging structure and direction information within subgraphs. CoMPILE \cite{mai2021communicative} innovates by enhancing message interactions and efficiently processing asymmetric relations, showcasing significant advances over traditional models.

\begin{figure*}[htbp]
    \centering
    \includegraphics[width=1\linewidth]{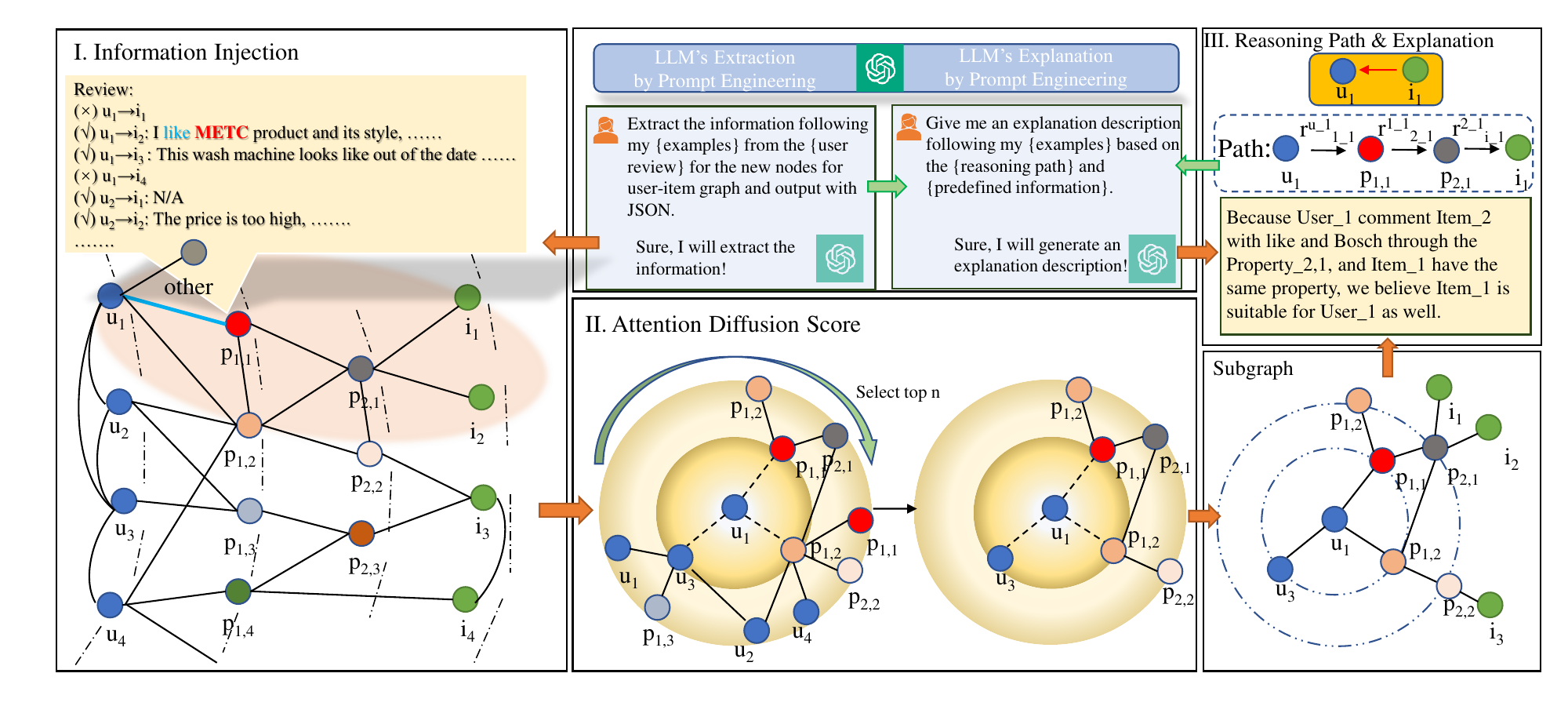}
    \caption{Framework of \method: I. Text information is extracted by the LLM and injected into the original knowledge graph by pre-define rules \{u is user, p is the property, i is item, r is relation\}; II. The attention score is calculated by the neighbours in different diffusion layers and subgraph could be generated. III. The final recommendation score is computed by the similarity function between the user and item, where a explanation path can be generated by LLM in this component.}
    \label{fig:framework}
\end{figure*}

\section{Method}

In this section, we introduce our model \method which is shown in \ref{fig:framework}.

\subsection{Preliminary}

In recommendation systems, a knowledge graph is formally defined as $\mathcal{G}$, $\mathcal{G}=\{ (e_h, r, e_t)|e_h \in \mathcal{E}, {e_t} \in \mathcal{E}, r \in \mathcal{R} \}$, where $\mathcal{E}$ is the sets of entities and $\mathcal{R}$ is the sets of relations. In this paper, we consider a special type of knowledge graph for recommendation system, denoted by $\mathcal{G}_\text{R}$. It contains a subset of a \textit{User} entities $\mathcal{U}$, a subset of \textit{Item} entities $\mathcal{I}$ and \textit{Properties} $\mathcal{P}$(item information, user portrait and so on), where $\mathcal{U}\cup\mathcal{I}\cup\mathcal{P}\subseteq\mathcal{E}$ and $\mathcal{U}\cap\mathcal{I}\cap \mathcal{P}=\varnothing$. These three kinds of entities are connected through relations $r_s$.
We give a relaxed definition of knowledge graph information injection as follows.

\textit{Definition 1 - Knowledge Graph Reconstruction:}
Knowledge graph reconstruction is the systematic assimilation of insights from textual data into a knowledge graph utilizing a LLMs. The LLM parses unstructured text, extracting entities including key words and semantic information, and discerning their relations to form a new $\mathcal{G}_\text{R}$. This mechanism, represented by a function $f: \mathcal{T} \rightarrow \mathcal{G}_\text{R}$, transcribes the all kinds of predefined targets from the text into graph structures—triples of the form $\{(\acute{e_h}, \acute{r}, \acute{e_t})\}$—thus enriching the existing KG with new, verifiable information.

\textit{Example:} Consider the comment text "I like \METC's wash machine colour." The LLM identifies \textit{user}, \textit{\METC}, \textit{wash machine} and \textit{colour}, and relations \textit{like} and \textit{belong}, forming the $\mathcal{G}_\text{R}$ triples: \textit{user} $\xrightarrow{\text{like}}$ \textit{wash machine} $\xrightarrow{\text{belong}}$ \textit{\METC}.

\textit{Definition 2 - Sub-Graph Reasoning:} 
In the reconstructed KGs, subgraph reasoning is defined as a sequence of diffusion from step $s$ to step $s+1$ staring from the central entity $e_{s,i}$, denoted by $d_{{s+1},i}(e_{s,i}, \cdots, e_{s,j})=\Theta \{e_{s,0}  \xleftrightarrow{r_{s,0}} e_{s,1}, e_{s,0} \xleftrightarrow{r_{s,1}} e_{s,2}, \cdots, e_{s,i} 
\xleftrightarrow{r_{s,k}} e_{s,j} \}$ , where $s \in S$ in the s-th step of attention-based diffusion process, $\Theta$ donates a function to select $N$ entities as the central entities for the $s+1$ step, the $i \in N $ is the $i$-th central entity in step $s$, especially when $s=0$, $i$ is the number of the users, $k$ is the $K$-th relations between the $e_{s,i}$ and $e_{s,j}$, and $j \in J$ is the $j$-th neighbour of central entity $e_{s,i}$ .

\textit{Definition 3 - Explainable Recommendation:} 
Explainable Recommendation is defined as given a sequence path $\mathcal{SP} = \{e_{0,i} \xleftrightarrow{r_{0,i}} e_{1,i} \xleftrightarrow{r_{1,i}} \cdots \xleftrightarrow{r_{s-1,k}} e_{s,i} \}$ generated by the subgraph, especially $\textit{User} \leftrightarrow \textit{Item}$ in recommendation system, the goal is to find a recommendation score function $S(\textit{User}, \textit{Item})$ considering the reasoning path $\mathcal{SP}$. And the recommendation result, reasoning path, and the predefined information will be fed to the LLMs to generate a explanation description.

\subsection{KG Reconstruction and Explanation Generation}

\subsubsection{\textbf{LLMs for Information Extraction and Injection}}
Our method commences with the decomposition of information using a LLM. This process involves the extraction of new entities and relationships from a given text by predefined targets. The prompt engineering technique is employed to refine queries which guide the LLM towards precise extractions. Newly identified entities and relations are then integrated into the existing KG, resulting in a reconstructed KG that encapsulates the enriched data. There are two examples for entity and relation extraction respectively.

\begin{minipage}{0.95\columnwidth}
    \centering
    \begin{tcolorbox}
        \small
        \textbf{Prompt Example for Entity Extraction:} From the review "\texttt{<Review>}", pinpoint significant dates, the events or action as relations. I will provide you some examples.
    \end{tcolorbox}
    \vspace{1mm}
\end{minipage}

This prompt facilitates the extraction of time entities and their associated events from user reviews (\texttt{<Review>}).

\begin{minipage}{0.95\columnwidth}
    \centering
    \begin{tcolorbox}
        \small
        \textbf{Prompt Example for Relation Extraction:} For the review "\texttt{<Review>}", identify any sentiments expressed. Output \textbf{<Positive>} for positive sentiments and \textbf{<Negative>} for negative sentiments. I will provide you some examples.
    \end{tcolorbox}
    \vspace{1mm}
\end{minipage}

The above prompt guides the LLM in analyzing reviews (\texttt{<Review>}) and structuring the output as a sentiment-indicative targets.

Specific rules are established to govern the integration of new information into the KG. These rules are tailored by product analyst to the scenario at hand, encompassing relations such as emotions, preferences, and quantifiable attributes (e.g., price, color, style), as well as time-connected entities such as significant dates. All extracted information will form a new link and embed it into the existing KG according to the customized set of rules.

\subsubsection{\textbf{LLMs for Post-hoc Explanation Generation}}

In the stage of generating interpretable descriptions, we provide the large language model with meaningful contextual prompts, including predefined key targets and subgraphs or paths generated by the subgraph reasoning process. Then, according to the requirements, we generate a segment of language description. Specific instances can be referenced in Table \ref{tab:case1} and Table\ref{tab:case2}. Below is the template we need for generation.

\begin{minipage}{0.95\columnwidth}
    \centering
    \begin{tcolorbox}
        \small
        \textbf{Prompt Example for Explainable Description:} Generate an explanation for this recommendation "\texttt{<item->user>}", based on the predefined target: "\texttt{<targets>}", and the reasoning path"\texttt{<path> }". I will provide you some answering examples.
    \end{tcolorbox}
    \vspace{1mm}
\end{minipage}

This prompt facilitates to generate the final explanation description based on the customized information.




\subsection{\textbf{Attentive Diffusion Subgraph Reasoning}}

We introduce an attention-based mechanism to construct a entity-centric subgraph, harnessing the entity's interaction history and contextual relevance within the knowledge graph. Initially, the subgraph centers around the user, progressively expanding by assimilating nodes based on their attention scores which signify their contextual importance to the entity.

At any given diffusion step $s$, we commence with the current entity subgraph $\mathcal{G}_{s,e}$ and identify the set of newly added nodes, designated as neighbor nodes $\mathcal{N}_{{s},e}$. Specifically, the initial subgraph $\mathcal{G}_{0,u}$ contains solely the user node, with neighbours $\mathcal{N}_{0,u}$. The next phase involves computing the attention scores for the edges $E_{s,e}$ connected to $\mathcal{N}_{s,e}$. This set of $E_{s,e}$ encompasses relationships between the central entity $\mathcal{G}_{s,e}$ and their $\mathcal{N}_{{s},e}$. The attention mechanism then evaluates the importance of these edges, determining the significance of the $E_{s,e}$ to the $\mathcal{G}_{0,u}$. 

Consequently, entity scores are calculated based on the aggregated attention scores of their associated edges, selecting the top $N$ nodes $N(\mathcal{N}_{{s},e})$ from the entire set of one-hop neighbors, rather than $N$ nodes per neighbor node. These top-scored nodes are then appended to the subgraph, diffusing it to $g_{s+1,e}$.

In the step $s$, specifically, the attention module is responsible for assigning scores to edges based on their relevance to the entity embedding. In this stage, the attention score $\alpha$ of an edge $(e_i,r,e_j)$ is computed by the following equations


\begin{equation}
    \acute{\alpha}(e_i,r,e_j) = \theta \left((W_2\delta(W_1(h_{e_u}\Vert h_{e_i})))\cdot h_{e_j}\right)
\end{equation}

\begin{equation}
    \alpha (e_i,r,e_j) = \frac{e^{\acute{\alpha}(e_i,r,e_j)}}{\sum_{(e_i,r,e_j) \in E_{s,e}}e^{\acute{\alpha}(e_i,r,e_j)}},
\end{equation}
where $\theta = \text{Sigmoid}$, $\delta=\text{LeakyReLU}$, $W_1 \in$ $\mathbb{R}^{ {d_1}X 2d}$, $W_2 \in$ $\mathbb{R}^{ d X {d_1}}$, ${d_1}$ is a parameter that controls the size of the trainable matrices $W_1$ and $W_2$, $h_{e_u} \in$ $\mathbb{R}^{d}$ and $h_{e_u}$ represents the emedding of user entities $u$, $h_{e_i}, h_{e_j} \in$ $\mathbb{R}^{d}$ and $h_{e_i}, h_{e_j}$ represents the embedding of entities $e$, $(a \Vert b)$  denotes concatenation of embedding vectors $a$ and $b$, $\alpha (e_i,r,e_j)$ represents the attention score of edge $(e_i,r,e_j)$ at step $s$. The edge $(e_i,r,e_j)$ is from the set of edges $E_{s,e}$.

After obtaining the attention scores of the edges, $e_j$ entity scores are derived by aggregating the attention scores from the edges to their corresponding entities.

\begin{equation}
    \widetilde{\text{score}}(e_j)=\sum_{(e_i,r,e_j) \in E_{s,e}} {\text{sore}}(e_i)\cdot \alpha(e_i,r,e_j)
\end{equation}

\begin{equation}
    \text{score}(e_j)=\frac{e^{\widetilde{\text{score}}(e_j)}}{\sum_{j \in J}e^{\widetilde{\text{score}}(e_j)}},
\end{equation}
where score($e$) represents the attention score of entity and $score(e_i)$ has been calculated in the last run. Then s-th step user subgraph is updated by adding the m highest scoring nodes to the user subgraph. 

After selecting the top $N (N<=J)$ score neighbours from the all neighbours $e_j$,  the $N(\mathcal{N}_{{s},e})$ need to be re-scored by normalization. Additionally, $N$ is a hyperparameter used to control the subgraph size. $v_n \in V_n$ will become the new central entity in step $s+1$.

\begin{equation}
    v = \text{score}(e_n) = \frac{e^{\widetilde{\text{score}}(e_n)}}{\sum_{n \in N}e^{\widetilde{\text{score}}(e_n)}}
\end{equation}

\subsection{\textbf{Recommendation Scoring}}

After constructing the user subgraph through the aforementioned attention-based scoring diffusion method, the next stage is to compute the recommendation scores for item nodes within this Sub-Knowledge-Graph that are connected to this user subgraph.

The recommendation scoring process, visualized in \ref{fig:framework}, moves beyond linear reasoning by encoding the relationships between different paths within the user subgraph. This allows the model to capture a comprehensive representation of user interests and to reflect the interplay of these paths when generating recommendations.

Initially, we delineate the subgraph pertinent to the item node from the overarching user subgraph. This subgraph is then encoded to extract user preferences using a subgraph encoder. The encoding process is formulated as follows:

\begin{equation}
    h_{g_{s}} = \sum_{i \in N} h_{v_{s, i}}, s \in \{1,2,...\}
\end{equation}

\begin{equation}
    h_{u, \{g_{1}, g_{2}\}} = W_4\delta(W_3(h_{u}\Vert h_{g_{1}}\Vert h_{g_{2}})),
\end{equation}
where $h_{v} \in$ $\mathbb{R}^{d}$ and 
$h_{g_{s}}$ represents the embedding of subgraph $g$ including $v_{s,i}$ in step $s$.
The set $V_{u_k}$ comprises the lists of nodes at a distance $k$ from the user node within the subgraph. $W_3 \in$ $\mathbb{R}^{\hat{d_2} X 3d}$, $W_4 \in$ $\mathbb{R}^{ d X \hat{d_2}}$,
$\hat{d_2}$ is a parameter that controls the size of the trainable matrices $W_3$ and $W_4$. The user $u$ subgraph is composed of nodes user $u$, $g_{1}$ and $g_{2}$, so $h_{u, \{g_{1}, g_{2}\}} \in \mathbb{R}^{d}$ represents the embedding of the user $u$ subgraph.

Subsequently, the similarity score between this subgraph representation and the item's embedding is calculated:

\begin{equation}
\text{sim}(u,it_m) = \theta(h_{u, \{g_{1}, g_{2}\}} \cdot h_{it_{m}}),
\end{equation}
where sim($u,it_m$) is the similarity score between user $u$ and the m-th item $it_m$.

To account for the connection between the subgraph and the full user subgraph, we integrate the score derived during subgraph construction as the weight for each subgraph. This weighted score is computed as follows:


\begin{equation}
 \text{S}_{u,it_m} = \sum_{e_j \in g_{3}} v \cdot \text{sim}(u,it_m).
\end{equation}

Here, S$_{u,it_m}$ denotes the final recommendation score for an item $it_m$ relative to user $u$, incorporating both the encoded subgraph-user-item relationship and the subgraph's relevance within the user subgraph.

The existing objective function only guides the generation of the subgraph and does not indicate which subgraph truly reflects the user's preferences. Therefore, we introduce a specific optimization objective for the recommendation module, defined as:

\begin{equation}
    L_{u} = -\frac{1}{|Y(u)|} \sum_{u_i \in Y(u)}\log(\text{S}_{u,it_m}).
\end{equation}

This objective function minimizes the negative log-likelihood over the set of items positively associated with the user $u$, denoted by $Y(u)$, refining the model's ability to deduce user preferences from subgraphs.

\section{Experiments}
In this section, our experiment is mainly divided into two parts. The first part is a performance experiment, which mainly reflects the recommended performance of our subgraph-based inference by comparing it with other baselines. In the second part, we use a case study to explore the contribution of our model's LLM information injection to the entire recommendation system, and tests the improvement ability of subgraph (SG) and parameter sensitivity of our design module. In the experiments of this work, we used entity representations pre-trained via TransE-based in ReMR.

\begin{table}[ht]
    \centering
    \caption{Statistics of the datasets.}
    \label{tab:statistics}
    \begin{tabular}{c c c c c}
        \toprule
          &\METC &Beauty&Cell Phones & Clothing \\
         \midrule
         \#User & 1004 & 22,363 & 27,879 &39,387 \\
         \#Items & 1017 & 12,101 &  10,429 &23,033 \\
         \#Entities & 2482 & 224,080 &  163,255 & 425,534\\
         \#Relations & 12 & 16 & 16 & 16 \\
         \#Interactions & 3636  & 198.58K & 194.32K & 278.86K \\
         \#Triples & 129980  & 37.73M &  37.01M & 36.37M \\
         \bottomrule
    \end{tabular}
\end{table}

\begin{table*}[h]
\centering

\caption{Performance on top-10 recommendation between the baselines and our model. The results are computed in the test set and are given as percentages \%. The best baseline results are underlined.}
\label{tab: main result}
\begin{tabular}{c cccc cccc cccc cccc}
\hline
      & \multicolumn{4}{c}{Beauty} & \multicolumn{4}{c}{Cell Phones} & \multicolumn{4}{c}{Clothing}  \\  
      Measures(\%)     & NDCG   & Recall & HR     & Prec. & NDCG        & Recall & HR & Prec. & NDCG     & Recall & HR & Prec. \\
\hline
BPR          & 2.805& 5.032& 8.933& 1.173& 1.995& 3.534& 5.424& 0.623& 0.665 & 1.219  & 1.932 & 0.323 \\
DKN          & 1.923& 2.591& 8.812& 1.135& 1.672& 3.313& 4.580& 0.349& 0.375 & 0.724 & 1.492 & 0.119 \\
CKE          & 3.824& 6.241& 11.132& 1.422& 3.849& 6.981& 10.633& 1.073& 1.656 & 2.604 & 4.329 & 0.390 \\
KGAT         & 5.020& 7.794& 12.496& 1.535& 4.803& 7.982& 11.241& 1.134& 2.824 & 4.674 & 6.993 & 0.603 \\
PGPR         & 5.489& 8.324& 14.347& 1.692& 4.921& 8.383& 11.832& 1.280& 2.863 & 4.797 & 7.024 & 0.719 \\
ReMR       & \underline{5.878} & \underline{8.982 }& \underline{15.606}& \underline{1.906} & \underline{5.294} & \underline{8.724} & \underline{12.498} & \underline{1.337}  & \underline{2.977} & \underline{5.110} & \underline{7.426} & \underline{0.766} \\
\method        & \textbf{6.187}  & \textbf{9.788} & \textbf{16.103} & \textbf{1.928} & \textbf{5.755} & \textbf{9.753} & \textbf{13.378} & \textbf{1.433} & \textbf{3.520} & \textbf{6.050} & \textbf{8.739} & \textbf{0.922} \\
\hline
Improvement(\%)  & +5.257 & +8.974 & +3.185 & +1.154 & +8.708 & +11.795 & +7.041 & +7.180 & +17.450 & +18.395  & +17.681 & +20.366 \\
\hline
\end{tabular}
\end{table*}

\subsection{Experiment Settings}

\subsubsection{\textbf{Datasets}}

We conducted experiments on three datasets, among them are the Amazon review dataset collated in KGAT\cite{KGAT19}, and \METC dataset. \METC dataset is a collection of \METC order data from different channels.  
We used three datasets from the Amazon dataset, namely Cell Phones, Beauty and Clothing. When dealing with the Amazon dataset, we followed the data processing methods in PGPR\cite{PGPR19}. 
The details of the dataset statistics are shown in Table \ref{tab:statistics}.

\subsubsection{\textbf{Baselines}}
We consider six recommendation approaches as baselines in the following experiments. These baselines are divided into three categories, Matrix Factorization-based models, KG embedding models and path reasoning models.

BPR\cite{bpr12}: BPR is a personalized ranking algorithm based only on user product interaction information through Bayesian posterior optimization.

DKN\cite{dkn18}: This is a recommendation model that combines knowledge graph reality and convolutional neural network

CKE\cite{CKE16}: CKE uses TransR\cite{lin2015learning} to obtain semantic embeddings from the knowledge graph to enhance collabrative filtering.

KGAT\cite{KGAT19}: KGAT learns entity embeddings from knowledge graphs through graph attention networks combined with GNN and attention mechanisms.

PGPR\cite{PGPR19}: PGPR is a knowledge graph path reasoning model based on reinforcement learning.

ReMR\cite{ReMR22}: ReMR relies on its own data to obtain higher-order abstraction information for the entities in the knowledge graph to create a multi-layer knowledge graph.

\subsubsection{\textbf{Evaluation Criteria}}
For all approaches, we adopted four evaluation criteria to evaluate the top-5 recommendations of each user in the test set, including Normalized Discounted Cumulative Gain (NDCG), Recall, Hit Rate (HR), and Precision (Prec.).

\subsubsection{\textbf{Implementation Details}}
In our model, the entity embedding dimensionality is 100.The hyperparameter of subgraph size is set to a a maximum of 100. We train the parameters  with Adam optimization, batch size of 256, and a number of training epochs of 10. 

\subsection{Performance Experiments}

\subsubsection{\textbf{Performance Comparison}}

We show the top-10 recommendation performance of our proposed method compared to all baselines. And comparison with the model ReMR is not possible in \METC dataset because of the lack of higher-order abstract information of entities. The specific information is shown in Table \ref{tab: main result}.

Table \ref{tab: main result} shows that our method consistently outperforms all baselines in terms of recommendation accuracy. On average, our model improves NDCG, Recall, HR and Precision by 5.36\%, 9.33\%, 6.58\%, and 6.46\%. This demonstrates that subgraph reasoning with LLM can help better infer user interests and improve recommendation performance.

It is noted that the algorithms that focus on node representation learning are not very effective on the Amazon data set. Maybe Amazon’s knowledge graph contains more information and user interests are more complex, so it is not appropriate to only focus on node embedding. And \method has the greatest performance improvement in recall, which to a certain extent shows that \method can fully learn user interests.

\subsection{Case Study for \METC}

\subsubsection{\textbf{Brief Introduction}}
In order to illustrate the explanation in recommendation process, we do a case study in \METC private dataset to visualize the reasoning paths and the improvement by LLM. 
\METC's product families are all durable goods and in diverse shopping domains and channels. Based on the market analysis, more than half of \METC consumers tend to purchase more than two categories of products in more than two e-commercial channels. Hence, analysing the decision making path of consumers and providing an accurate and explainable recommendation is challenging but critical for \METC business growth.

The \METC dataset includes three key aspects. 1) User table: it includes the user attributes, e.g., id, profiles, reviews, and preferred channels. 2) Order log: it includes the details of each order, e.g., user id, product id, product properties, review of the order. 3) User activities: it includes the event tracing data in the online/offline channels such as the click and view.


The Table~\ref{tab:bosch result} shows that our proposed model outperforms the baselines in all four measures. 

\begin{table}
    \centering
    \caption{Results in \METC Dataset}
    \label{tab:bosch result}
    \begin{tabular}{c cccc}
        \hline
         & \multicolumn{4}{c}{\METC}\\
        Subgraph Size & NDCG & Recall & HR & Prec.\\
        \hline
        BPR  & 14.961 & 20.941 & 23.705 & 2.420 \\
        DKN  & 11.137& 18.034& 19.821& 2.042 \\
        CKE & 9.436 & 19.687 & 22.709 & 2.371\\ 
        KGAT & 14.707 & 21.847 & 23.606  & 2.410\\
        PGPR & \underline{17.767} & \underline{24.253} & \underline{27.191} & \underline{2.799}\\
        ReMR & - & - & -  & -\\
        \method & 18.144 & 26.006 & 29.781 & 3.108\\
        Improvement(\%)  & +2.122& +7.228& +9.525&  +11.040\\
        \hline
    \end{tabular}
\end{table}

\subsubsection{\textbf{Explanation Visualization}}

\begin{figure}
     \centering
     \subfigure[Case study 1]{
     \includegraphics[width=\linewidth]{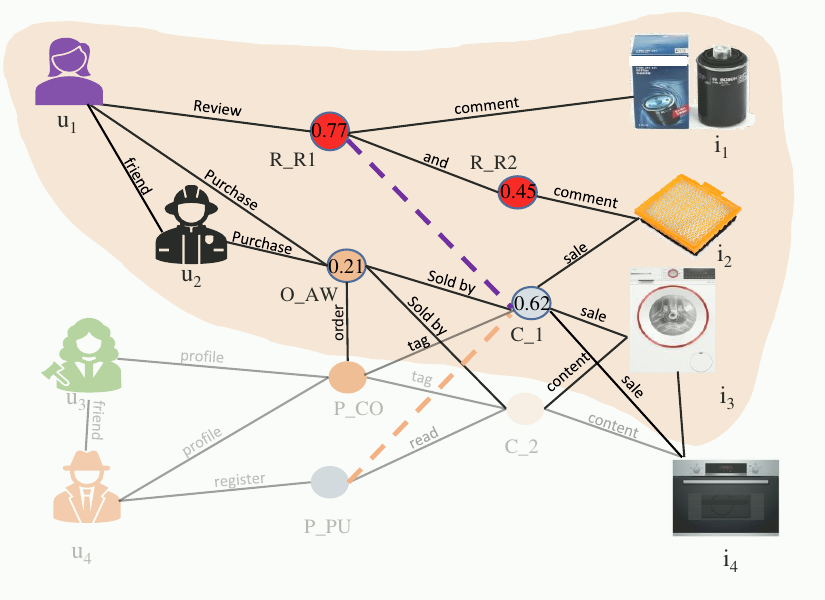}
     \label{fig:case1}}

     \subfigure[Case study 2]{
     \includegraphics[width=\linewidth]{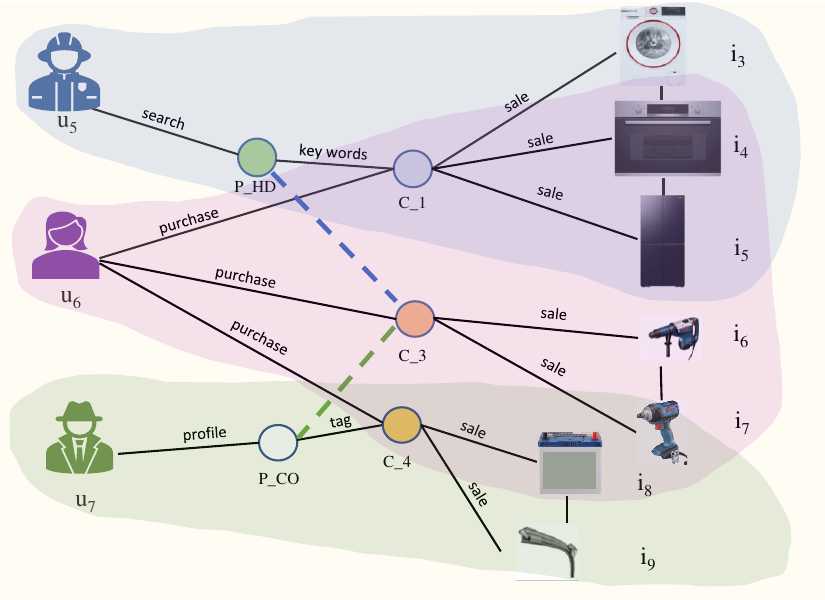}
     
     \label{fig:case2}}
     
     \caption{Real cases discovered by our model, each containing a subgraph which end nodes is predicted item by recommendation model. u1 – u4 : \METC users, i1 – i4 : Auto air filter related products, R\_R1: “review: reliable”, R\_R2 : “review: no smell”, O\_AW: “order: auto wiper”, S\_HD: “search: Home Decoration ”, C\_1: “channel: Channel 1 auto parts flagship store”, P\_CO: “profile: car owner” , C\_2: “channel: Channel 2”, P\_PU: “profile: premium user”, C\_3: “channel: Channel 3 ”, C\_4: “channel: Channel 4”.}
\label{fig:case}
\end{figure}

Figure~\ref{fig:case1} illustrates how our model show an explainable result and echo the business questions in the \METC application. It can be seen that for auto air filter users, the property like car owners and \METC premium user(user profiles), Channel 1 and Channel 2 (selling channels like E-commerce platforms), reliable and no smell (reviews), and auto wiper purchasing (cross-selling pair) are the main contributors in the recommendations, which indicates how the majority of users made their decisions. The red nodes are generated by the LLM and the blue dashed line is one of the most probable reasoning path.
Finally, the item$_4$-oven is highly probable to recommend to the user$_1$.(PS: the number in different nodes is the score for user$_1$).

In Figure~\ref{fig:case2}, it is another typical scenario. User$_5$ has only bought \METC home appliances on Channel 1. User$_6$ bought different \METC prodution in different channels such as \METC home appliances on Channel 1, \METC power tools on Channel 3, and \METC car accessories on Channel 4. User$_7$ only bought \METC car accessories on Channel 4. We finally recommended \METC cordless drill to user$_5$ and user$_6$. Explanation path comparison between path-based method and ours has been shown in Table \ref{tab:case1} and Table \ref{tab:case2}.

\subsubsection{\textbf{Observations and Analysis}}

In cases exemplified by example1 and example2, although the target items in the list, while the path-based method can invariably generate an existed path to suggest a target product to a specific user, where the paths seem far-fetched and even make no sense. This  is named as one type of "recommendation hallucination". Furthermore, other items in the recommendation list  for user that exist similar explainable paths, which can not reflect the requirement of users. 

In example3 and example4, the path-based recommendation approach indiscriminately suggests all products associated with User6 to both User5 and User7. For instance, car accessories are recommended to User5, and home appliances to User7. This strategy can lead to what may be described as another "recommendation hallucination" where the rationale provided by the system does not substantiate the recommendations made, thus potentially exerting a detrimental impact on decision-making processes.

In contrast, utilizing the reliable explanation paths offered by our method suggests that the recommendation system gains a more profound understanding of the relationships between users and items. Consequently, the items that appear in the recommendation list are more accurately aligned with the user's needs.

\begin{table}
\centering
\caption{Ablation study}
\label{tab:ablation}
\begin{tabular}{c cccc cccc}
\hline
    &\multicolumn{4}{c}{\METC} \\

Measures(\%)  & NDCG   & Recall & HR     & Prec.\\
\hline
w/o Review      & 16.378 & 23.045 & 25.863 & 2.878 \\
\hline
w/o SG   & 15.964 & 21.582 & 22.266 & 2.246 \\
\hline
Full model      & 18.144 & 26.006 & 29.781 & 3.108 \\
\hline
\end{tabular}
\end{table}

\begin{table}
    \centering
    \caption{Influence of Subgraph Size}
    \label{tab:prameter}
    \begin{tabular}{c cccc}
        \hline
         & \multicolumn{4}{c}{\METC}\\
        Subgraph Size & NDCG & Recall & HR & Prec.\\
        \hline
        60  & 17.258 & 23.433 & 26.494 & 2.739 \\
        80  & 18.144 & 26.006 & 29.781 & 3.108 \\
        100 & 17.584 & 24.750 & 28.586 & 2.928\\ 
        \hline
    \end{tabular}
\end{table}

\begin{table*}[th]
\centering
\caption{Case 1 of \textit{explanation} comparison by Path-based reasoning and \method.}
\vspace{-3mm}
\label{tab:case1}
\resizebox{2\columnwidth}{!}{
\begin{tabular}{c|ccc}
\toprule
Example 1 & \multicolumn{3}{c}
{\METC Oven($Item_4$) $\to$  $User_1$ }                                                                    \\  \midrule
Path-based reasoning 
& \multicolumn{1}{|p{2in}@{}|}
{$User_1$ $\to$ \underline{purchase} $\to$ \textit{auto wiper} $\to$ \underline{sold by} $\to$ \textit{C\_1}  $\to$  \underline{sale} $\to$ $Item_4$ } 
& \multicolumn{1}{c|}{Explanation} 
& \multicolumn{1}{p{3in}@{}}
{Because the user purchased a auto wiper directly from \METC and Channel 1 platform sale the Channel 2 production including auto wiper, wash machine, oven and so on, system recommends oven to the user.}\\ 
\midrule
Subgraph-based reasoning by ours                   
& \multicolumn{1}{|p{2in}@{}|}
{$User_1$ $\to$ \underline{review} $\to$  \textit{reliable}  $\to$ \textit{C\_1}  $\to$  \underline{sale} $\to$ $Item_4$}         
& \multicolumn{1}{c|}{Explanation} 
& \multicolumn{1}{p{3in}@{}}
{User commented the \METC heating system with reliable and no smell. system guesses that user pay more attention to "reliable". Channel 1 is a reliable platform which sale number of \METC production including auto wiper, wash machine, oven and so on. Thus system recommends oven to the user.}\\
\midrule

Example 2 & \multicolumn{3}{c}{\METC Heating System ($Item_3$) $\to$  $User_4$ }            \\  \midrule
Path-based reasoning 
& \multicolumn{1}{|p{2in}@{}|}
{$User_4$ $\to$ \underline{profile} $\to$  \textit{car owner}   $\to$ \underline{tag} $\to$ \textit{C\_1}  $\to$ \underline{sale} $\to$ $Item_3$}  
& \multicolumn{1}{c|}{Explanation} 
& \multicolumn{1}{p{3in}@{}}
{User was labeled car owner as his one of profile and Channel 1 platform has the same tag, and Channel 1 sale number of \METC production including auto wiper, wash machine, oven and so on so the system recommends wash machine to this user.}\\ 
\midrule
Subgraph-based reasoning by ours                   
& \multicolumn{1}{|p{2in}@{}|}
{$User_4$ $\to$ \underline{register} $\to$  \textit{premium}   $\to$ \underline{read} $\to$ \textit{Wechat}  $\to$ \underline{content} $\to$ $Item_3$}    
& \multicolumn{1}{c|}{Explanation} 
& \multicolumn{1}{p{3in}@{}}
{This user is a premium of \METC in Wechat and read \METC washing machine article as well. system guesses that this user have high probable willing to buy \METC wash machine. Also, Channel is one of the biggest sale platform who sale number of \METC production including auto wiper, wash machine, oven and so on so the system recommends wash machine to this user.}\\

\midrule

\bottomrule
\end{tabular}}
\vspace{-3mm}
\end{table*}

\begin{table*}[th]
\centering
\caption{Case 2 of \textit{explanation} comparison by Path-based reasoning and \method.}
\vspace{-3mm}
\label{tab:case2}
\resizebox{2\columnwidth}{!}{
\begin{tabular}{c|ccc}
\toprule
Example 3 & \multicolumn{3}{c}
{\METC Cordless Drill ($Item_7$) $\to$  $User_5$ }                                                                    \\  \midrule
Path-based reasoning 
& \multicolumn{1}{|p{2in}@{}|}{$User_5$ $\to$ \underline{search} $\to$ \textit{Home Decoration} $\to$ \underline{key words} $\to$ \textit{C\_1}  $\to$  \underline{purchase} $\to$ $User6$ $\to$ \underline{purchase} $\to$ \textit{C\_3} $\to$ \underline{sale} $\to$ $Item_7$ } 
& \multicolumn{1}{c|}{Explanation} 
& \multicolumn{1}{p{3in}@{}}{Because $User_5$ and $User_6$ bought \METC household appliances in Channel 1, system guesses what $User_6$ bought in Channel 3 is suitable for $User_5$. Thus, system recommends cordless drill to $User_5$ }\\ 
\midrule
Subgraph-based reasoning by ours                   
& \multicolumn{1}{|p{2in}@{}|}{$User_5$ $\to$ \underline{search} $\to$  \textit{Home decoration} $\to$ {\underline{key words}} $\to$\textit{C\_3} $\to$  \underline{sale} $\to$ $Item_7$}         
& \multicolumn{1}{c|}{Explanation} 
& \multicolumn{1}{p{3in}@{}}{$User_5$ searched home decoration as key words in Channel 1 and bought some household appliances, system guesses $User_5$ needs some tools to install the household appliances, so system recommends cordless drill to $User_5$} \\
\midrule

Example 4 & \multicolumn{3}{c}{\METC Cordless Drill ($Item_7$) $\to$  $User_7$ }                                                                    \\  \midrule
Path-based reasoning 
& \multicolumn{1}{|p{2in}@{}|}{$User_7$ $\to$ \underline{profile} $\to$  \textit{car owner}   $\to$ \underline{tag} $\to$ \textit{C\_4}  $\to$ \underline{purchase} $\to$ $User6$ $\to$ \underline{purchase} $\to$ \textit{C\_3} $\to$ \underline{sale} $\to$ $Item_6$}  
& \multicolumn{1}{c|}{Explanation} 
& \multicolumn{1}{p{3in}@{}}{Because $User_7$ and $User_6$ bought \METC car battery in Channel 4, we guess what $User_6$ bought in Channel 3 is suitable for $User_7$. Thus system recommends cordless drill to $User_7$}.\\ 
\midrule
Subgraph-based reasoning by ours                   
& \multicolumn{1}{|p{2in}@{}|}{$User_7$ $\to$ \underline{profile} $\to$  \textit{car owner}   $\to$ {\underline{tag}} $\to$ \textit{C\_3}  $\to$ \underline{sale} $\to$ $Item_6$}    
& \multicolumn{1}{c|}{Explanation} 
& \multicolumn{1}{p{3in}@{}}{$User_7$ is a car owner and bought some car accessories such as battery and auto wiper” in Channel 4, so system guesses $User_7$ need some power tools to install these accessories, so system recommends \METC cordless drill to $User_7$}\\

\bottomrule
\end{tabular}}
\vspace{-3mm}
\end{table*}

\subsubsection{\textbf{Ablation Studies}}

On the \METC dataset, We further study the importance of each module of our model in Table \ref{tab:ablation}, where w/o Review means that our remove the entities which are extracted from user review and w/o SG is our model that removes subgraph generating. The experimental results shows that, after removing these modules, the noise could not be attenuated and uncompleted information could interfere with reasoning process, leading to poor results, while the subgraph generation mechanism of our model can better capture user interests.


\subsubsection{\textbf{Parameters sensitivity}}

As can be seen in Table \ref{tab:prameter}, the hyperparameter of subgraph size indicates the maximum number of nodes for subgraph expansion in each round of subgraph reasoning. Larger hyperparameter means that we generate larger subgraphs with more nodes in them. Larger subgraphs tend to contain more information, which on the one hand increases the tolerance for the performance of the subgraph inference module, but on the other hand it increases the performance requirements for the subgraph scoring module.

\section{Conclusion}

In this paper, we present a novel integration of a LLM, and subgraph of KG generation to foster the development of an explainable recommendation system. This represents the inaugural effort to amalgamate these advanced technologies for enhancing recommendation systems. Specifically, the LLM is leveraged to distill key information from textual data, which is then systematically incorporated into the KG using pre-defined rules. Furthermore, we introduce an attention-based diffusion mechanism for the generation of subgraphs, facilitating the construction of nuanced user representations, which is used for calculating the recommendation score with the item profiles.
To evaluate the effectiveness of our proposed model, we conducted a series of experiments across three publicly available datasets and one proprietary dataset from \METC. The empirical results unequivocally demonstrate that our model outperforms existing state-of-the-art models across all datasets. Importantly, the case studies underscore our model's capacity to provide explicit, user-centric explanations for recommendations, effectively addressing specific user comments, relieving the "recommendation hallucination" effectively.

Moreover, our framework unveils substantial potential for further optimization in terms of efficiency and opens new avenues for research in recommendation systems employing Large Language Models. This underscores the promising intersection of advanced computational models and practical application domains, heralding a new era of intelligent, user-focused recommendation systems.

We hope that our work will provide some inspiration for methods in the same scenario, particularly in the integration of Large Language Models and Knowledge Graphs, and offer strategic insights for market analysts. For instance, in the decision-making process, because the system gives high score for some key words in review, marketing analyst could pre-define these key words for user review, thereby enhancing the system's robustness.

\newpage

\balance

\bibliographystyle{ACM-Reference-Format}
\bibliography{7Reference}

\end{document}